\documentclass[flushrt,tighten]{aastex}

\slugcomment{submitted to The Astrophysical Journal Letters}
\shorttitle{Hot Spots in SNR~1987A}
\shortauthors{Lawrence {\it et al.}}

\begin{document}
\title{On the Emergence and Discovery of Hot Spots in Supernova
Remnant 1987A \footnote{Based in part on archival observations
obtained with the NASA/ESA {\em Hubble Space Telescope}, obtained at
the Space Telescope Science Insitute, which is operated by AURA,
Inc.\ under NASA contract NAS5-26555.}} 

\author{Stephen~S.~Lawrence,\altaffilmark{2}
  Ben~E.~Sugerman,\altaffilmark{2}
  Patrice~Bouchet,\altaffilmark{3}
  Arlin~P.~S.~Crotts,\altaffilmark{2}
  Robert~Uglesich\altaffilmark{2} and
  Steve~Heathcote\altaffilmark{3}}
\altaffiltext{2}{Department of Astronomy, Columbia University,
  New York, NY 10027; lawrence@, ben@, arlin@, rru@astro.columbia.edu}
\altaffiltext{3}{Cerro Tololo Inter-American Observatory, Casilla 603,
  La Serena, Chile; pbouchet@, sheathcote@noao.edu}

\begin{abstract}
We present the discovery of several new regions of interaction between
the ejecta and equatorial circumstellar ring of SN1987A, an interaction
leading to a much expanded development of the supernova remnant.
We also trace the development of the first such ``hot spot,''
discovered in 1997, back to 1995.\ \ Later hot spots seem to have emerged
by early 1999.\ \ We
discuss mechanisms for the long delay between the first and later spots.
\end{abstract}


\section{INTRODUCTION}
The development of SN~1987A provides an unprecedented opportunity
to observe, at high spatial, spectral, and temporal resolution,
the birth of a supernova remnant (SNR).\ \ Observations
of SNR~1987A might reveal the angular and
velocity distribution of the ejecta by highlighting
where it impacts a presumably well-known nebular structure.
While SNR~1987A has been observed for some years in X-rays and radio,
only optical/near-IR data have sufficient sensitivity
and angular resolution to detail the structures presented here.

The early discovery of these interaction regions is crucial.
We demonstrate how they are best revealed with difference imaging,
in H$\alpha$ and
[\ion{N}{2}], and in \ion{He}{1}~1.083$\mu$m, for which
the newly-formed spots have greatest contrast above the circumstellar
equatorial ring (ER).\ \ We
detect several new interaction sites, and trace
known features back to epochs well before their actual discoveries.

\section{OBSERVATIONS AND REDUCTIONS}
As the contrast between the first hot spot and the underlying ER
was greater in \ion{He}{1}~1.083$\mu$m than in optical lines
at early times (\S 3),
we have conducted ground-based monitoring of the ejecta-ER
collision in this line.
Data were taken on the CTIO 4-m Telescope with Tip-Tilt
first order wavefront correction, on  
5 nights between 1997 November and 1998 October
[days 3926--4243 after SN core collapse]
with the CIRIM imager [1.75~h total integrations]
and 1999 December 25 [day 4688]
with the OSIRIS imager/spectrometer [3.5~h total integration].\ \ To  
accurately measure small changes in the ER and
hot spot fluxes, we apply image subtraction techniques
\citep{Tom96}
as brieflly outlined below.  We also deconvolve images from each
epoch, using multiscale maximum entropy
\citep{Pan96} 
to improve the FWHM to $\sim$0\farcs 3.

We also analyze data from the {\em HST} public archive, making use of:
a) STIS spectra (G750M grating) from 1997 April 26 [day 3715]
and 1998 November 14 [day 4283];
b) NICMOS images taken through F108N on 1997 December 9
[day 3942];
and c) WFPC2 images taken through F336W, F439W, F555W, F656N, F658N,
F675W, and F814W between 1994 February and
1999 April [days 2537--4440].\ \ Pipeline
calibrated STIS spectra were taken directly from the archive,
while the NICMOS dither pattern was manually
re-mosaiced.  For PC chip images, aberrant
pixels were replaced, and multiple exposures were co-added  with
cosmic-ray rejection.  Recent press-release images from
\citet{Kir00}\footnote{STScI-PR00-11 available at {\em http://oposite.stsci.edu/pubinfo/pr/2000/11}}
are highly deconvolved, an invasive procedure which is
sensitive to the input PSF and the stability of the pixel map.
Furthermore, 
subtraction of deconvolved images
assumes that constant sources are mapped identically in each epoch.
To study hot spot variability, we opted for the
non-iterative, non-invasive procedure of PSF-matched difference
imaging.  The PC chip has a PSF FWHM $\sim$1.7 pixels, below 
Nyquist critical sampling, so we first convolved images with a
circular gaussian of $\sigma=0.65$, smoothing to a final
${\rm FWHM}\ga 2.2$ pixels.
Registered, convolved frames were then
PSF matched, photometrically scaled, and differenced using
{\it difimphot} calls to IRAF routines (e.g.\ {\em immatch}).\ \ All
flux calibrations used standard {\em HST} methods.

\section{DISCUSSION}
We propose to denote 
known and future hot spots as, e.g.\ \mbox{HS~1-029},
where the first digit represents the order of
discovery and the trailing three digits encode
the first reported imaging PA of the feature.  This scheme preserves the order
of appearance and provides a rough spatial location.
Names based soley on
discovery order will become confusing as new
knots appear at a higher rate.  The PA encoding
is approximate; different passbands and data quality
will naturally show small PA differences.

{\it Discovery of New Hot Spot Activity:}
The highest quality image (Figure~1a) from our ground-based \ion{He}{1}
monitoring (1998 October 6 with a ${\rm FWHM}=0\farcs 71$) shows
the ER and the first hot spot (\mbox{HS~1-029})
resolved between Stars 2 and 3.\ \ Figure~1b
displays a difference between 1999 and 1998 images.
While the continued brightening of \mbox{HS~1-029}
is obvious, a second spot (\mbox{HS~2-104}) has now appeared---the
first new locus of activity
\citep{Bou00}
since the
discovery of \mbox{HS~1-029} in 1997 April spectra
\citep{Pun97}.
Figure~1c presents a deconvolution
of the 1999 December image, with a final ${\rm FWHM}=0\farcs 34$.\ \ Several
regions of brightening are distributed around the ER.\ \ Such
low-level deconvolution features require confirmation,
seen below with {\em HST} images.

Figure~1d shows a 1997 December 11 NICMOS F108N mosaic with
${\rm FWHM}=0\farcs 10$.\ \ Star~5
and \mbox{HS~1-029} are clearly resolved on the ER, with no
other hot spots significantly detected.
The ratio of F108N flux from \mbox{HS~1-029} to the total
ER+\mbox{HS 1-029} flux is 0.06$\pm$0.01.\ \ The
same ratio from 1998 February 6 WFPC2 imaging 
is 0.03$\pm$0.01 for both F656N and F675W, and is 0.02$\pm$0.01 for
F502N in 1997 July 10.\ \ Since F656N and F675W imaging followed
the F108N epoch by nearly two months while \mbox{HS~1-029}
was brightening rapidly, {\em at early
times} hot spots appear at greater contrast above
the ER in \ion{He}{1}~1.083~$\mu$m than in strong
optical transitions, as found in ground-based data
\citep{Kun98}.
We recommend diligent monitoring of
the ER in \ion{He}{1}, with adaptive
optics and with a resuscitated NICMOS, to detect
new hot spot activity promptly.

{\it Confirmation in Archival {\em HST} Data:}
As reported by
\citet{Law00},
a STIS G750M/6581\AA\ spectrum from
1998 November shows
the spectroscopic signature of a developing hot spot.  
The 0\farcs 2 wide slit was positioned at PA 102\fdg 9, and
includes \mbox{HS~2-104} and perhaps a fraction of \mbox{HS~4-091}.\ \ As
Figure~1e shows,
a wide-velocity feature, with FWHM $\sim$150~km~s$^{-1}$, is
redshifted from the ER at this position by 
$\sim$50~km~s$^{-1}$, correcting for the displacement between the slit
center and \mbox{HS~2-104}.\ \ This feature's flux is 
$\sim$2$\times10^{-16}$~erg~s$^{-1}$~cm$^{-2}$ in H$\alpha$.\ \ This
is similar in description to \mbox{HS~1-029} on the near side of the ER,
seen in 1997 April by
\citet{Son98},
blueshifted
over the range 0-250~km~s$^{-1}$.\ \ With a measured H$\alpha$ flux of
\mbox{$\sim$1.0$\times10^{-15}$}~erg~s$^{-1}$~cm$^{-2}$,
\mbox{HS~1-029} on day 3715 appears more developed than \mbox{HS~2-104} on
day 4283.

Following our ground-based discovery of new spot activity, \mbox{HS~2-104}
and four other new spots were promptly
detected in STIS F28X50LP and WFPC2 F656N imaging
\citep{Mar00,Gar00}.
As \mbox{HS~2-104} was detectable in archival {\em HST}
spectra, we reviewed the public STIS, NICMOS, and WFPC2 data
to search for the earliest appearances of all spots.
No spots other than \mbox{HS~1-029} and \mbox{HS~2-104} were seen
in the STIS or NICMOS archives, and \mbox{HS~2-104} appeared only
in the STIS spectra from 1998 November described above.
Figure~1f displays the F656N difference image between
1999 April and 1998 February.
All of the five ``new'' spots are detected,
along with at least three more sites of increasing
F656N flux along the ER inner edge.
All recently discovered activity began {\em no later than
1999 January} [day 4337].\ \ The
positions, fluxes, and earliest detections of nine spots,
measured in the F656N 1999$-$1998 difference image relative to the
centroid of SN~1987A in F656N in 1994 February, are presented
in Table~1
($\pm3\degr$ in PA, $\pm$0\farcs 03 in $r$, and
$\pm$10\% in flux for the first four, $\pm$30\% for the rest).\ \ The
first eight spots are also detected in 1999-1998 F675W differences.
There are also marginal detections of new spots at ($166\degr$, 0\farcs 51)
and ($183\degr$, 0\farcs 48), but these are faint and seen in only one epoch.

Known and unknown transient hot pixels contaminate roughly 1\% of the PC
chip, and since most of the WFPC2 imaging of SN1987A was not spatially
dithered within filter sets, these pixels present a persistant, stochastic
and non-gaussian source of noise, to which both difference imaging and
deconvolution analyses are precariously sensitive.
A full presentation of our analysis technique
will be presented in a subsequent paper.  Briefly, we cosmetically
correct known bad pixels, while tracking their redistribution through the
%%%convolution and geometric-registration processes.  To minimize false
convolution and registration processes.  To minimize false
detections, a potential hotspot (residual in a difference image) must
not lie within a bad-pixel domain and must appear in at least two
difference images between mutually-exclusive data.
The dashed arrow in Figure~1f indicates a ``false'' spot caused
by a bad pixel.
Newly detected
spots should be considered tentative until they have been confirmed
with further imaging or spectroscopic data, e.g.\ \mbox{HS~7-289}
is confirmed in STIS spectra from 2000 May 1
\citep{LSA00}.
The remaining unconfirmed spots could also be reverse-shock
H$\alpha$ emission at low velocity
\citep{Son98,Mic98}.

{\it The Evolution of \mbox{HS~1-029}:}
Figure~2 shows difference images for seven
WFPC2 bands spanning optical wavelengths.
Each row portrays a single filter,
and each column the differences between a given year and
1994.\ \ Orientation is indicated by the inset graphic.
The 1994 images have been scaled down to match the fading ER,
in order to highlight flux changes on this structure.
As such, residuals from field stars remain, notably for Star~5.\ \ The
evolution of \mbox{HS~1-029} is clearly demonstrated.  Although
first reported in STIS spectroscopy from 1997 April, 
\mbox{HS~1-029} is clearly detectable in 1996 February in F502N, F555W,
F675W, and F658N, and is seen faintly
in F555W and F675W in 1995 March.
The spot is evident in 1996 even in differences without PSF matching.
\mbox{HS~1-029} began
developing {\em on or before 1995 March} [day 2933].

A large arc of transient flux brightens after 1994 in F675W, F656N,
and F658N.\ \ First noted by
\citet{Gar96},
it spans $180\degr$$<$PA$<$$235\degr$ in F675W and nearly encircles
the ER in F658N in 1996.\ \ Relative
fluxes in these filters in 1996 and 1997 suggest it is mostly
[\ion{N}{2}] emission.
This could be limb-brightened
recombination near the base of the bipolar lobes, where the walls of
the ``hourglass'' join with the ER
\citep{CKH95}.
A lower
density implies a long delay before maximum intensity,
consistent with line-emission models of
\citet{Lun91}, 
and a peak in 1998 indicates a density $\sim$10 times lower
than in the ER.\ \ Continued
monitoring in [\ion{N}{2}]
could reveal additional hourglass structure.

\section{CONCLUSIONS}
\mbox{HS~1-029} began within 8 years after the SN and at least
3--4 years ahead of the subsequent spots.
Did this asymmetry arise in the velocity
of the ejecta, in the density of the blue supergiant (BSG) wind, or in the
morphology of the ER itself?
The BSG wind should have been azimuthally smoothed by rotation of
the progenitor, and relatively undisturbed by the ER
(interior to the reverse shock $\sim$5$\times10^{17}$~cm from the
star).\ \ Rayleigh-Taylor instabilities are expected in this decelerating wind,
however over the $\sim$20000~yr age of the ER
\citep{CH91}
the sound speed would have allowed smoothing of these 
over scales of tenths of a parsec.
This should be modelled over timescales for the BSG wind propagation
to the ER ($\sim$1000~yr).\ \ If
the ER/ejecta interaction asymmetry were due to a
large-scale misalignment between a bipolar ejecta pattern and the
ER axis, a counter-jet should have impacted opposite \mbox{HS 1-029},
contrary to observation.
It seems likely that the asymmetry is due to structures in the ER
or ejecta velocity field with fairly narrow extent in position angle,
i.e.\ thin fingers or jets.
The homogeneity of the ER's shape versus the long delays between
the \mbox{HS~1-029} and later spots points to the ejecta
velocity field as the major cause.

Valuable opportunities to obtain diagnostic spectra of
the hot spots at very early times have been lost.
If the fainter spots presented here are confirmed,
then the ejecta have reached the ER over a wide
range of PAs, and soon little pristine ER material will remain.
Frequent ground and space-based
monitoring of this rapidly developing event is required,
with emphasis on difference imaging in
\ion{He}{1}~1.083$\mu$m, H$\alpha$, and [\ion{N}{2}].

\clearpage

\figcaption[Figure1revis.eps]
{(a) A \ion{He}{1}~1.083$\mu$m ground-based image of SNR~1987A from
1998 October.
(b) A \ion{He}{1} difference image between 1999 December and 1998 October.
New activity centered on \mbox{HS~2-104} is indicated.
(c) A deconvolution of the 1999 December \ion{He}{1} image.
(d) A NICMOS F108N mosaic from 1997 December.
Frames a--d are 6\arcsec\ square with north up, east to the left.
(e) A STIS 52$X$0.2 G750M spectrum centered at 6581\AA,
5\farcs 0 wide and 36\AA\ tall with slit at
PA $= 283\degr$.\ \ H$\alpha$
from \mbox{HS~2-104} is indicated.
(f) A difference between WFPC2 F656N images from 1999 April and
1998 February,
3\farcs 0$\times$2\farcs 0 with north up, east to left.
Radial lines indicate the PAs of the hot spots, which are located
slightly outwards from the inner segment.
The dashed arrow flags a bad pixel residual.
\label{fig1}}

\figcaption[Figure2revis.eps]
{An array of WFPC2 difference images with 1994 subtracted from 1995--1999.
Each row presents a particular filter,
noted to the left; the first frame in the F656N row is
F658N.\ \ Each column shows a given year, subtracted by 1994,
noted at top.
Each frame is 2\farcs 1$\times$2\farcs 3 with north indicated by the
graphic.
The 1994 images have been scaled down to match the mean ER emission;
regions fading relative to 1994 appear black.
Note that \mbox{HS~1-029} began as early as 1995 March in F555W
and F675W.
\label{fig2}}

\clearpage

\begin{deluxetable}{lccccc}
\footnotesize
\tablecaption{Hot Spots in SNR 1987A. \label{tbl-1}}
\tablewidth{0pt}
\tablehead{
\colhead{Spot} & \colhead{PA} &
\colhead{$r$} &
\colhead{$\Delta f_{99-98}$\tablenotemark{a}} &
\colhead{$t_{earliest}$\tablenotemark{b}} & \colhead{Refs.} \\
\colhead{} & \colhead{(\degr E of N)} &
\colhead{(\arcsec)} &
\colhead{(10$^{-17}$~erg} &
\colhead{(days)} & \colhead{} \\
\colhead{} & \colhead{} &
\colhead{} &
\colhead{cm$^{-2}$~s$^{-1}$)} &
\colhead{} {}\\
}
\startdata
1-029  & 27 & 0.56 & 425 & 2933 & 1 \\
2-104  & 106 & 0.68 & 62.1 & 4283 & 2 \\
3-126  & 123 & 0.60 & 58.1 & 4337 & 3 \\
4-091  & 91 & 0.68 & 30.8 & 4337 & 4 \\
5-139  & 139 & 0.54 & 9.5 & 4337 & 4 \\
6-229  & 230 & 0.67 & 1.2 & 4440 & 4 \\
7-289  & 289 & 0.64 & 2.3 & 4440 & 5 \\
8-064  & 64 & 0.61 & 9.3 & 4440 & 6 \\
9-075  & 75 & 0.64 & 6.8 & 4440 & 6 \\
\enddata
\tablenotetext{a}{Flux difference in F656N between 1999 Apr
  and 1998 Feb, scaled so that fading ER emission cancels.}
\tablenotetext{b}{Earliest detection in {\em HST} data, days after SN.}
\tablerefs{(1) Pun et al.\ 1997; (2) Bouchet et al.\ 2000;
  (3) Maran et al.\ 2000; (4) Garnavich et al.\ 2000;
  (5) Lawrence et al.\ 2000; (6) first reported here.}
\end{deluxetable}

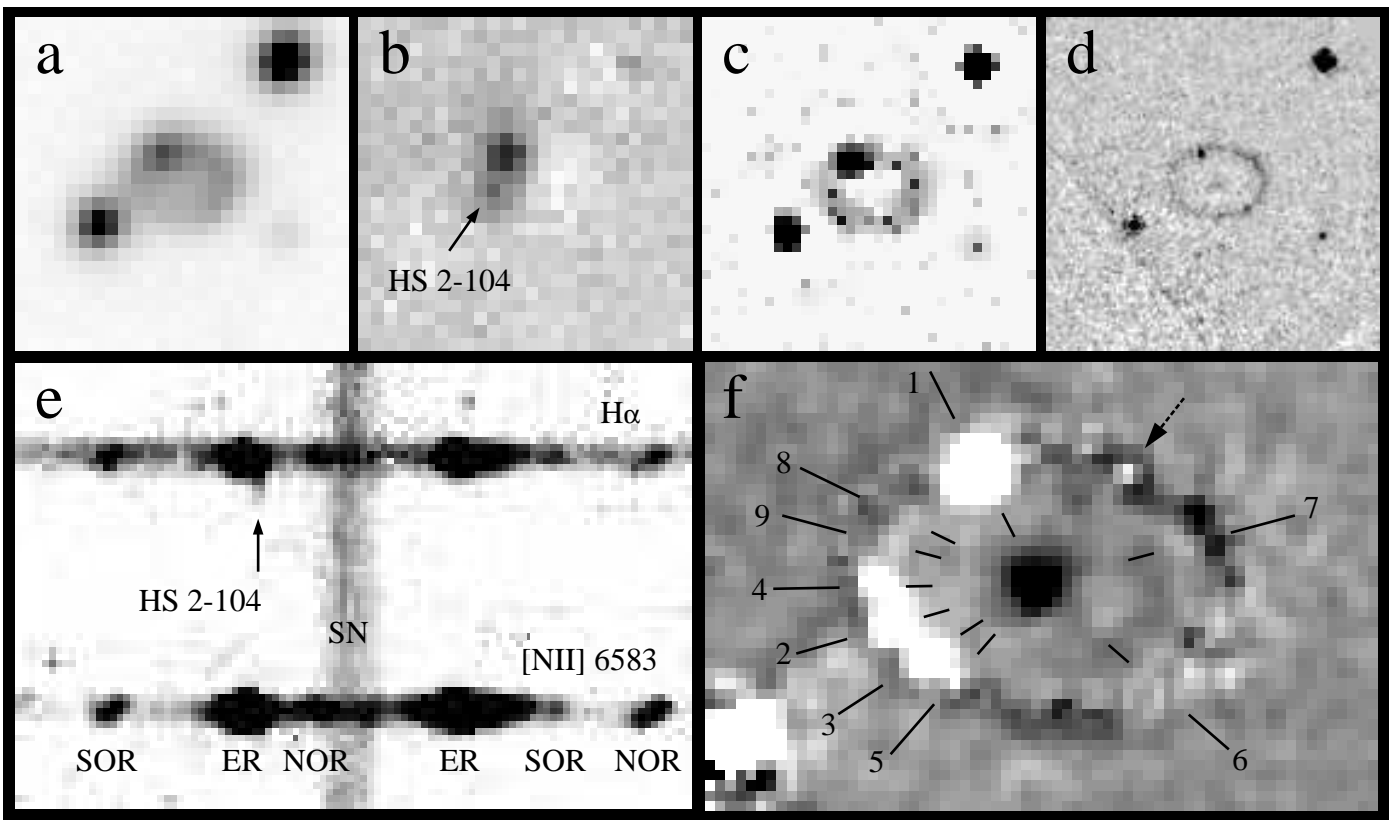
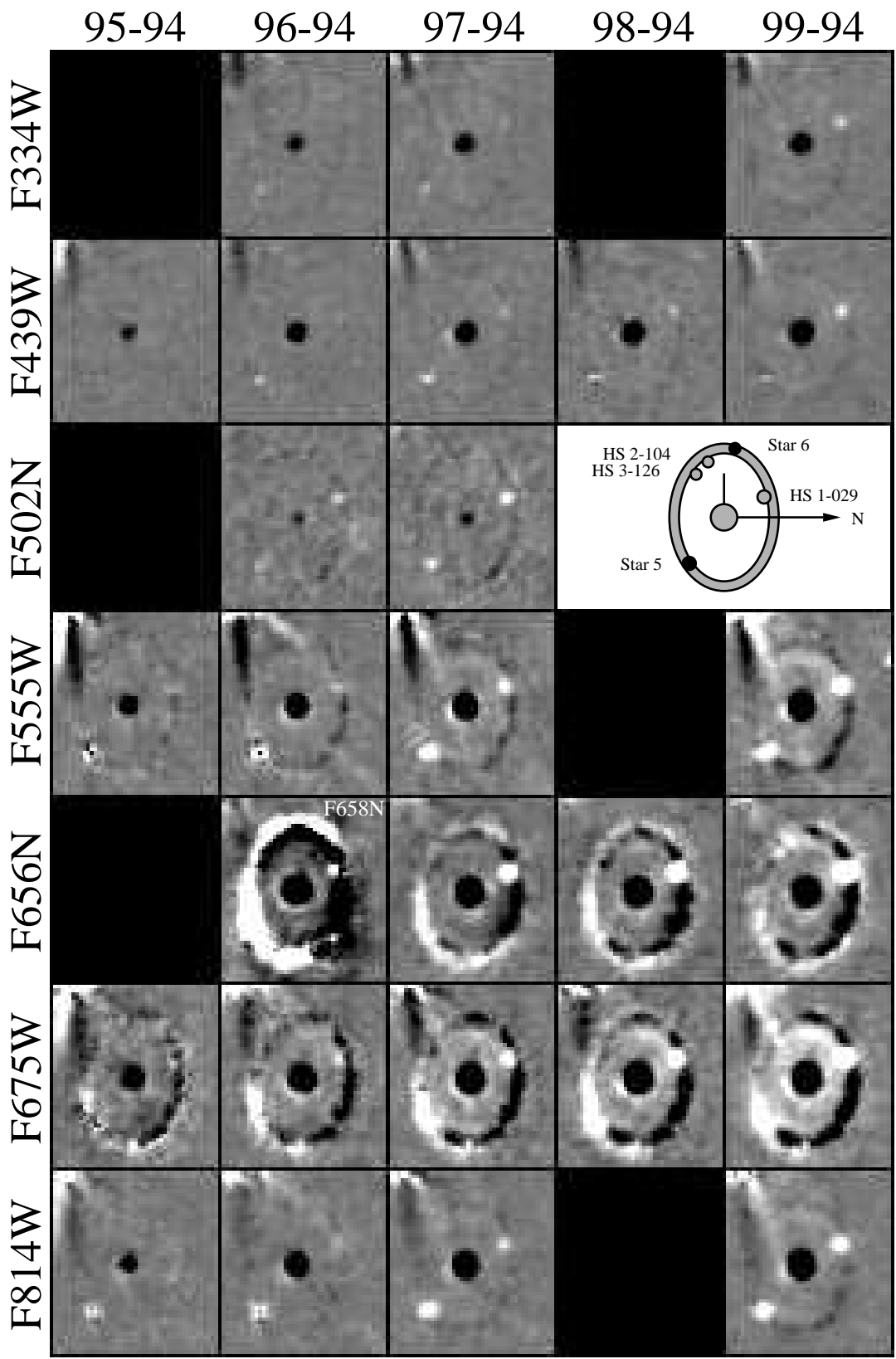
\end{document}